\documentclass[preprint,prd,nofootinbib,tightenlines,amsmath]{revtex4}
\usepackage{graphicx,epsfig}
\begin{document}

\title{\bf Glueball Production via Gluonic Penguin B Decays}

\vspace{1cm}

\renewcommand{\thefootnote}{\fnsymbol{footnote}}

\author{
Xiao-Gang He$^{1}$
\footnote{Email address: {\sf hexg@phys.ntu.edu.tw}}
and Tzu-Chiang Yuan$^{2}$
\footnote{Email address: {\sf tcyuan@phys.nthu.edu.tw}} }
\affiliation
{1.~Department~of~Physics~and~Center~for~Theoretical~Sciences,
~National~Taiwan~University,~Taipei,~Taiwan~10764,~R.O.C.\\
2.~Department~of~Physics,~National~Tsing~Hua~University,~Hsinchu,~Taiwan~300,~R.O.C.}

\renewcommand{\thefootnote}{\arabic{footnote}}

\date{\today}

\vspace{2cm}

\begin{abstract}
We study glueball $G$ production in gluonic penguin decay $B\to
X_s G$, using next-to-leading order $b\to s g^*$ gluonic penguin
interaction and an effective coupling of a glueball to two gluons.
The effective coupling allows us to study the decay rate of a
glueball to two pseudoscalars in the framework of chiral
perturbation theory. Identifying the $f_0(1710)$ to be a scalar
glueball, we then determine the effective coupling strength with
the branching ratio of $f_0(1710) \to K \overline K$. We find
that the charm penguin to be important and obtain a sizable
branching ratio for ${\rm Br}(B\to X_s G)$ in the range of $(0.7 \sim
1.7) \times 10^{-4}$. Rare hadronic $B$ decay data from B{\tiny A}B{\tiny AR}
and Belle can provide important information about glueballs.
\end{abstract}

\maketitle

The existence of glueballs is a natural prediction of QCD.
However, glueball state has not been confirmed experimentally. The
prediction for the glueball masses is a difficult task.
Theoretical calculations indicate that the lowest lying glueball
state is a scalar with a mass in the range of 1.6 to 2 GeV. Recent
quenched lattice calculations give a glueball mass $m_G$ equals $1710\pm 50 \pm
80$ MeV \cite{lattice}. These results support that the scalar
meson $f_0(1710)$ to be a glueball. Phenomenologically,
$f_0(1710)$ could be an impure glueball since it can be
contaminated by possible mixings with the quark-antiquark states
that have total isospin zero
\cite{lee-weingarten,burakovsky-page,giacosa-etal,close-zhao,he-li-liu-zeng,cheng-chua-liu, fariborz}. These mixing
effects can be either small \cite{giacosa-etal,close-zhao,he-li-liu-zeng, fariborz} or
large \cite{lee-weingarten,burakovsky-page,giacosa-etal,cheng-chua-liu},
depend largely on the mixing schemes
one chose to do the fits and complicate the analysis. For
simplicity, we will ignore these mixing effects and assume
$f_0(1710)$ is indeed a glueball throughout this work.

\begin{figure}[hbt]
\begin{center}
\includegraphics[width=8cm]{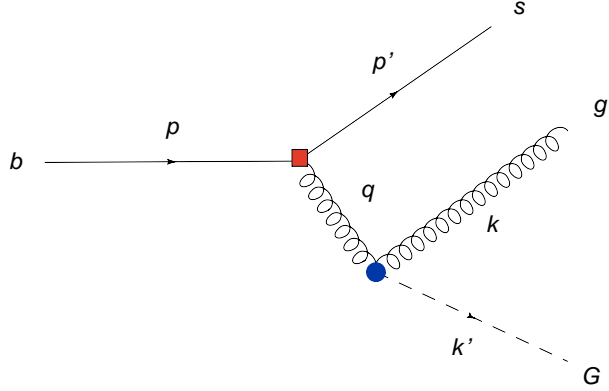}
\end{center}
\caption{The leading diagram for $B\to X_s G$.} \label{Feynman-1}
\end{figure}

Since the leading Fock space of a glueball $G$ is made up of two
gluons, production of glueball is therefore most efficient at a
gluonic rich environment like $J/\psi$ or $\Upsilon$  $\rightarrow
(gg)\gamma \rightarrow G\gamma$ \cite{shifman,jpsi}. Direct
glueball production is also possible at the $e^+e^-$ \cite{brodsky}
and hadron \cite{cky} colliders.
In this work, we point out another interesting mechanism to detect
a glueball via the rare inclusive process $B\to X_s G$ decay. The
leading contribution for this process is shown in Fig.
[\ref{Feynman-1}], where the squared vertex refers to the gluonic
penguin interaction and the round vertex stands for an effective
coupling between a glueball and the gluons.
The gluonic penguin $b\to s g^*$ has been studied extensively in
the literature and was used in the context for inclusive
decay $b \to s g \eta'$ \cite{hou-tseng}. The effective
interaction for $b\to s g^*$ with next-to-leading QCD correction
can be written as \cite{he-lin}
\begin{eqnarray}
\Gamma_{\mu a} =- {G_F\over \sqrt{2}}{g_s\over 4 \pi^2}
V_{ts}^*V_{tb} \bar s(p') \, [ \Delta F_1(q^2 \gamma_\mu - q_\mu \not\!
q) L - i m_b F_2  \sigma_{\mu\nu} q^\nu R ] \, T^a \, b(p),
\label{penguin}
\end{eqnarray}
where $\Delta F_1 = 4\pi(C_4(q, \mu) + C_6(q,
\mu))/\alpha_s(\mu)$ and $F_2 = -2 C_8(\mu)$ with $C_i(q,\mu)$  $(i=4,6, \, {\rm and} \, 8$) the
Wilson's coefficients of the corresponding operators in the
$\Delta B = 1$ effective weak Hamiltonian, $q = p - p'=k+k'$, and
$T^a$ is the generator for the color group. We will use the
next-to-leading order numerical values of $\Delta F_1$ and $F_2$ \cite{he-lin}. The top
quark contribution gives $\Delta F^{\rm top}_1 = -4.86$ and $F^{\rm top}_2 = +0.288$
at $\mu = 5$ GeV; whereas the charm quark contribution involves a $q^2$ dependence
through $C_4^{\rm charm}(q,\mu) = C^{\rm charm}_6(q,\mu) = P^{\rm charm}_s(q,\mu)$ with
\begin{eqnarray}
P^{\rm charm}_s(q,\mu) & = & {\alpha_s(\mu) \over 8 \pi} C_2(\mu) \left (
{10\over 9} + Q(q,m_c,\mu) \right) \; ,
\end{eqnarray}
and
\begin{eqnarray}
Q(q,m,\mu) & = &  4\int^1_0 dx \, x\,(1-x) \ln \left[m^2 - x(1-x)q^2\over \mu^2 \right]
 \; \, \nonumber \\
& = &
\frac{2\left[ 3q^2\ln (m^2/\mu^2)-12m^2-5q^2 \right]}{9q^2} +
\frac{4(2m^2+q^2)\sqrt{4m^2-q^2}}{3\sqrt[3]{q^2}}{\rm arctan}\sqrt{\frac{q^2}{4m^2-q^2}} \; .
\nonumber\\
\end{eqnarray}
Here $m_c = $ 1.4 GeV is the charm quark mass and $C_2(\mu = 5\,\mbox{GeV}) = 1.150$.

The following effective coupling between a scalar glueball and two
gluons was advocated recently by Chanowitz \cite{chanowitz}
\begin{eqnarray}
{\cal L} = f {\cal G} G^a_{\mu\nu}G^{a\mu\nu},\label{chano}
\end{eqnarray}
where ${\cal G}$ is the interpolating field for the glueball $G$,
$G^a_{\mu\nu}$ is the gluon field strength, and $f$ is an unknown
coupling constant. This form of effective coupling suggests that
glueball couples to the QCD trace anomaly.

The interaction of a scalar glueball with light hadrons through the
trace anomaly can be
formulated systematically by using techniques of chiral Lagrangian,
as was briefly mentioned in \cite{chm}.
The kinetic energy and the symmetry breaking mass terms for the light
pseudoscalar mesons are given by \cite{donoghue}
\begin{eqnarray}
{\cal L}_\chi = {f^2_\pi \over 8} [ {\rm Tr}(\partial^\mu \Sigma
\partial_\mu \Sigma^\dagger) +  {\rm Tr}(\xi^\dagger \chi \xi^\dagger + \xi \chi
\xi)],
\end{eqnarray}
where  $f_\pi =   132$ MeV being the pion decay constant,
$\xi^2 =\Sigma = \exp (2i\Pi/f_\pi)$ with $\Pi$ the $SU(3)$
pseudoscalar octet meson,
\begin{eqnarray}
\Pi = \left ( \begin{array}{cccc}
{1 \over \sqrt{2}}\pi^0+{1 \over \sqrt{6}}\eta  &\pi^+ &K^+\\
\pi^-&- {1 \over \sqrt{2}}\pi^0 +{1 \over \sqrt{6}}\eta&K^0\\
K^-&{\overline K}^0 &-{2 \over \sqrt{6}} \eta \end{array}\right ) \;,
\end{eqnarray}
and
\begin{eqnarray}
\chi= 2B_0 \, {\rm diag}(\hat m, \hat m, m_s) = {\rm diag}(m^2_\pi, m^2_\pi,
 2m^2_K - m^2_\pi)
 \end{eqnarray}
with $B_0 = 2031$ MeV.
Here we have neglected the isospin breaking effects due to small mass
difference between the light $u$ and $d$ quarks and used $\hat m = m_u = m_d$.
The QCD trace anomaly is well known and given by \cite{donoghue}
\begin{eqnarray}
\Theta^\mu_\mu
& = & - \frac{b\alpha_s}{8\pi} G^a_{\mu\nu}G^{a\mu\nu} + \sum_q m_q \bar q q \; ,
\end{eqnarray}
where $b = 11 - 2 n_f / 3$ is the QCD one-loop beta function with $n_f = 3$ being the number of light quarks.
Treating the effective interaction (\ref{chano}) as a perturbation to the energy
momentum stress tensor, one would then modify $\Theta^{\mu}_{\mu}$ to be
\begin{eqnarray}
-\frac{b\alpha_s}{8\pi} G^a_{\mu\nu}G^{a\mu\nu} \left(1+ f \frac{8\pi }{ b \alpha_s} {\cal G} \right) +
\sum_q \left( m_q +f_q {\cal G} \right) \bar q q  \; ,
\end{eqnarray}
where $f_q = f\alpha_s
m_q(16\pi\sqrt{2}/3\beta)\ln\left[(1+\beta)/(1-\beta)\right]$ being the one-loop induced $G q \bar q$
coupling \cite{chanowitz} with $\beta =
(1-4m^2_q/m_G^2)^{1/2}$. Note that $f_q$ is proportional to $f$.
The corresponding chiral Lagrangian is thus modified accordingly
\begin{eqnarray}
{\cal L}_\chi &=&{1\over 8} f^2_\pi \left(1+  f {8\pi\over b \alpha_s} {\cal G} \right)
{\rm Tr} \left[ \partial^\mu \Sigma \partial_\mu \Sigma^\dagger \right] \nonumber\\
&+&{1\over 8} f^2_\pi \, {\rm Tr} \left[ \xi^\dagger (\chi+2B_0 f_\chi
{\cal G})\xi^\dagger + \xi (\chi + 2 B_0 f_\chi {\cal G}) \xi \right],
\label{trace}
\end{eqnarray}
where $f_\chi = {\rm diag}(f_u, f_d, f_s)$. Using the above chiral Lagrangian, one can then calculate
the decay rates for $G \rightarrow \pi^+\pi^-,\pi^0\pi^0,K^+K^-,K^0{\overline K}^0,\eta^0\eta^0,$ and
$\pi^0\eta^0$.
Since $f_0(1710)$ is interpreted as the glueball, one can use the
experimental branching ratio ${\rm Br}(f_0(1710) \rightarrow K
{\overline K}) = 0.38^{+0.09}_{-0.19}$ and its total width $137\pm
8$ MeV \cite{pdg} to be our input. Together with the decay rate
formulas derived from the chiral Lagrangian (\ref{trace}) and a value of the
strong coupling constant $\alpha_s = 0.35$ extracted from the experimental data
of $\tau$ decay, one can estimate the unknown coupling
$f =0.07^{+0.009}_{-0.018}$ ${\rm GeV}^{-1}$.

In our estimation of $f$ given above, we have extrapolated low
energy theorems to the glueball mass scale. One might overestimate
the hadronic matrix elements in due course. This implies that the
extracted value of $f$ would be too small.

Decay $G \to K\overline K$ has also been obtained using perturbative
QCD calculations  \cite{chm}. This approach can also give some estimate
of the amplitude. The problem facing this approach is that the
energy scale may not be high enough to have the perturbative QCD contribution
to dominate. Using the asymptotic light-cone wave functions, we
find that for a given branching ratio for $G \to K\overline K$ decay,
the resulting $f$ would be about 20 times larger than the value
obtained above using the chiral approach. Incidentally, the value
of $f$ derived from the chiral Lagrangian is within a factor of 2
compared with the value estimated just by using the free quark
decay rate of $G \rightarrow s \bar s$ \cite{cky}. In our later
calculations, we will use the conservative value
$f=0.07^{+0.009}_{-0.018}$ ${\rm GeV}^{-1}$ determined using the
chiral Lagrangian given in Eq.(\ref{trace}). This will give the
most conservative estimate for the branching ratio since the
chiral approach gives the smallest $f$.

With the two effective couplings given in Eqs.(\ref{penguin}) and (\ref{chano}),
the following decay
rate for $b \rightarrow sgG$ (Fig.(\ref{Feynman-1})) can be obtained readily
\begin{eqnarray}
\Gamma_{b \rightarrow sgG}
& = & \left( {N_c^2 - 1 \over 4 N_c} \right) {G_F^2 m_b^5 \vert V^*_{ts} V_{tb} \vert^2 \over 2^5 \pi^3}
\left( {g_s \over 4 \pi^2} \right)^2 (m_b f)^2 \nonumber\\
& \;\; & \times \int_0^{(1-\sqrt{x'})^2} dx \int_{y_-}^{y_+} dy \left\{
|\Delta F_1|^2 c_0 + {\rm\bf Re}(\Delta F_1 F^*_2) {c_1\over y} +
|F_2|^2 {c_2\over y^2} \right\} \; ,
\label{dis}
\end{eqnarray}
with
\begin{eqnarray}
y_{\pm} = {1 \over 2} \left[ (1-x+x') \pm \sqrt{(1-x+x')^2 - 4 x'} \right] \; .
\end{eqnarray}
In Eq.(\ref{dis}), $N_c$ is the number of color
and $c_{0,1,2}$  are given by
\begin{eqnarray}
c_0 & = & \frac{1}{2} [-2x^2 y + (1-y)(y-x')(2x+y-x')] \; ,\nonumber\\
c_1 & = & (1-y)(y-x')^2 \; ,\nonumber\\
c_2 & = & \frac{1}{2} [2x^2y^2 -(1-y)(y-x')(2xy-y+x')] \; ,
\end{eqnarray}
with $x = (p' + k)^2/m^2_b$,  $y=(k+k')^2/m^2_b$, and
$x'=m^2_G/m^2_b$.

\begin{figure}[hbt]
\begin{center}
\includegraphics[width=10cm]{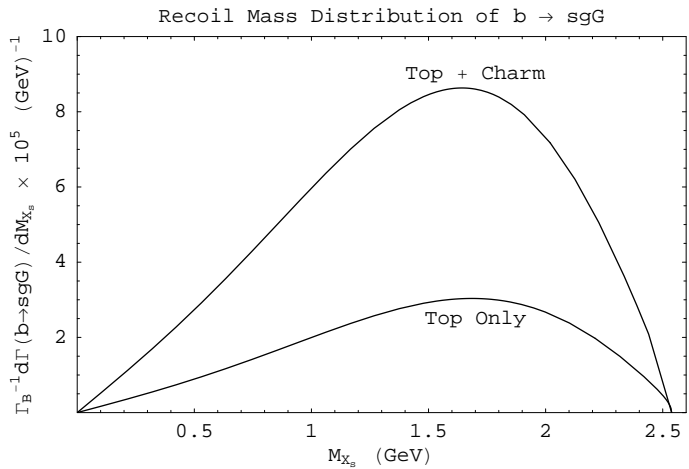}
\end{center}
\caption{$d{\rm Br}(B\to X_s G)/dM_{X_s}$ in unit of $10^{-5}$. In
the figure the $b$ quark mass $m_b$ is taken to be 4.248 GeV,
$\alpha_s = 0.21$ at the $b$ quark scale, $\tau_B = 1.674 \times
10^{-12}$ s, and $f = 0.07$ GeV$^{-1}$.} \label{Feynman-2}
\end{figure}

Using the value of $f$ determined above from the chiral
Lagrangian, we find the branching ratio for $b\to s  g G$ to
be $4.5\times 10^{-5}(f \cdot \mbox{GeV}/0.07)^2$ with just the leading
top penguin contribution to $\Delta F_1$ is taken into account. The correction from the
charm penguin is nevertheless substantial and should not be neglected. Inclusion of both top and charm
penguins gives rise to an enhancement about a factor of 3 in the branching ratio
${\rm Br}(b \to s gG) \approx
1.3\times10^{-4}(f \cdot \mbox{GeV}/0.07)^2$. Since $f_0(1710)$ has a
large branching ratio into $K\overline K$, the signal of scalar
glueball can be identified by looking at the secondary $K
\overline K$ invariant mass. The recoil mass spectrum of $X_s$ can
also be used to extract information. The distribution of $d{\rm
Br}(b \to s g G)/dM_{X_s}$ as a function of the recoil mass of
$X_s$ is plotted in Fig.(\ref{Feynman-2}).

Analogously one can also study inclusive $B$ decays into a
pseudoscalar glueball $\tilde G$, whose leading effective coupling to the gluon can be
parameterized as \cite{cornwall-soni},
\begin{eqnarray}
{\cal L} = \tilde f \tilde {\cal G} \tilde G^a_{\mu\nu}
G^{a\mu\nu} \; \;\;\; {\rm with} \;\;\; \; \tilde G^a_{\mu\nu} =
{1 \over 2}\epsilon_{\mu\nu\alpha\beta} G^{a\alpha\beta}  \; ,
\end{eqnarray}
and $\tilde{\cal G}$ denoting the interpolating field for the
pseudoscalar glueball. The decay rate for this case can be deduced
from the scalar glueball one by replacing the coupling $f$ with $\tilde f $ and
the mass $m_G$ with the pseudoscalar glueball mass $m_{\tilde G}$ in Eq.(\ref{dis}). This also reproduces previous
result obtained in Ref.\cite{hou-tseng} for a similar process
$b \rightarrow sg\eta^\prime$. We therefore expect similar
distribution and branching ratio for the pseudoscalar glueball
production from the $B$ decay as in the scalar case that we have
studied in this work.

Recently, BES has observed an enhanced decay in $J/\psi \to \gamma
\eta' \pi\pi$ with a peak around the invariant mass of
$\eta'\pi\pi$ at 1835 MeV \cite{bes2}. Proposal has been made to
interpret this state $X(1835)$ to be due to a pseudoscalar
glueball \cite{hhh}. Taking $X(1835)$ to be a pseudoscalar
glueball, we would obtain a branching ratio of $3.7\times
10^{-5}(\tilde f \cdot  \mbox{GeV} / 0.07)^2$ with top penguin
contribution only, and is enhanced to $1.1 \times 10^{-4}(\tilde
f \cdot \mbox{GeV}/0.07)^2$ if charm penguin is also included. If the
coupling $\tilde f$ is of the same order of magnitude as $f$, the
branching ratio for $B\to X_s \tilde G$ is also sizable.
Pseudoscalar glueball may also be discovered in rare $B$ decays.
Since a pseudoscalar glueball cannot decay into two pseudoscalar
mesons, the identification of a pseudoscalar glueball necessitates
the study of the three-body system $\eta'\pi\pi$. This makes the
analysis more difficult compared with the case of a scalar
glueball.

To conclude, we have studied inclusive production of a scalar
glueball in rare $B$ decay through the gluonic penguin and an
effective glueball-gluon interaction. The branching ratio is found
to be of the order $10^{-4}$. $B$ decays into a pseudoscalar
glueball through gluonic penguin is also expected to be sizable.
Note that we have used a conservative estimate of $f$. Observation
of $B \to X_s f_0(1710)$ at a branching ratio of order $10^{-4}$
or larger will provide an strong indication that $f_0(1710)$ is
mainly a scalar glueball. With more than 600 millions of
$B\overline B$ accumulated at Belle and more than 300 millions at
B{\tiny A}B{\tiny AR}, test of ${\rm Br}(b \to s g G)$ at the
level of $10^{-4}$ is quite feasible. We strongly urge our
experimental colleagues to carry out such an analysis.


\newpage
\noindent \acknowledgments We are grateful to K.m. Cheung and  J. P. Ma for many useful
discussions. This research was supported in part by the National
Science Council of Taiwan R.~O.~C.\ under Grant Nos.\ NSC
94-2112-M-007-010- and NSC 94-2112-M-008-023-, and by the National
Center for Theoretical Sciences.

\vspace{1cm}


\end{document}